\PassOptionsToPackage{table,prologue}{xcolor}
\documentclass[sigconf,nonacm,authorversion]{acmart}
\AtBeginDocument{%
  }
\usepackage{booktabs}
\usepackage{multirow}
\usepackage{amsmath}
\usepackage{enumitem}
\usepackage{float}
\usepackage{caption}
\usepackage{graphicx}
\usepackage{listings}

\begin{document}
\title{\textsc{AssertFlip}: Reproducing Bugs via Inversion of LLM-Generated Passing Tests}

\author{Lara Khatib}
\orcid{0009-0004-4770-8900}
\affiliation{%
  \institution{University of Waterloo, Canada}
  % \streetaddress{P.O. Box 1212}
  % \city{Dublin}
  % \state{Waterloo}
   \country{}
  % \postcode{43017-6221}
}
\email{lara.khatib@uwaterloo.ca}
\author{Noble Saji Mathews}
\orcid{0000-0003-2266-8848}
\affiliation{%
  \institution{University of Waterloo, Canada}
  % \streetaddress{P.O. Box 1212}
  % \city{Dublin}
  % \state{Waterloo}
   \country{}
  % \postcode{43017-6221}
}
\email{noblesaji.mathews@uwaterloo.ca}

 \author{Meiyappan Nagappan}
 \orcid{0000-0003-4533-4728}
\affiliation{%
  \institution{University of Waterloo, Canada}
  % \streetaddress{P.O. Box 1212}
  % \city{Dublin}
  % \state{Waterloo}
   \country{}
  % \postcode{43017-6221}
}
\email{mei.nagappan@uwaterloo.ca}

\renewcommand{\shortauthors}{Khatib et al.}

\begin{abstract}
Bug reproduction is critical in the software debugging and repair process, yet the majority of bugs in open-source and industrial settings lack executable tests to reproduce them at the time they are reported, making diagnosis and resolution more difficult and time-consuming. To address this challenge, we introduce \textsc{AssertFlip}, a novel technique for automatically generating Bug Reproducible Tests (BRTs) using large language models (LLMs). Unlike existing methods that attempt direct generation of failing tests, \textsc{AssertFlip}\ first generates passing tests on the buggy behaviour and then inverts these tests to fail when the bug is present. We hypothesize that LLMs are better at writing passing tests than ones that crash or fail on purpose. Our results show that \textsc{AssertFlip}\ outperforms all known techniques in the leaderboard of SWT-Bench, a benchmark curated for BRTs. Specifically, \textsc{AssertFlip}\ achieves a fail-to-pass success rate of 43.6\% on the SWT-Bench-Verified subset.
\end{abstract}

% \begin{CCSXML}
% <ccs2012>
%    <concept>
%        <concept_id>10011007</concept_id>
%        <concept_desc>Software and its engineering</concept_desc>
%        <concept_significance>500</concept_significance>
%        </concept>
%    <concept>
%        <concept_id>10010147.10010178</concept_id>
%        <concept_desc>Computing methodologies~Artificial intelligence</concept_desc>
%        <concept_significance>300</concept_significance>
%        </concept>
%  </ccs2012>
% \end{CCSXML}

% \ccsdesc[500]{Software and its engineering}
% \ccsdesc[300]{Computing methodologies~Artificial intelligence}

% \keywords{Bug Reproduction, Automated Test Generation, Large Language Models}

\maketitle

\section{Introduction}\label{sec:introduction}

Bug reproduction is an essential first step in the software bug-fixing process \cite{chaparro2019assessing}, where software developers attempt to replicate the bug to observe the faulty behaviour and understand the root cause \cite{vyas2014bug}. When a bug is discovered, a report is written in natural language and submitted to a bug-tracking system, containing relevant information about the issue. These reports often include a detailed description of the issue, step-by-step reproducing instructions, observed vs. expected behaviour, software version details, and supporting materials such as screenshots and videos to help developers investigate the bug \cite{zimmermann2010makes}. The reproduction steps outlined in the bug report can be converted into Bug Reproducible Test (BRT): a test case that fails when the bug is present and passes once the bug is fixed \cite{cheng2025agentic, mundler2024swt}. Previous research has shown that developers often rely on BRTs to diagnose, debug, and fix bugs, in addition to ultimately verifying bug fixes \cite{beller2018dichotomy}. However, despite their importance, many studies show that BRTs are rarely written at the time of bug reporting in open-source projects. Instead, they are usually added after the fix to validate that the bug has been resolved and will not reoccur \cite{cheng2025agentic}. Mundler et al. \cite{mundler2024swt} found that in the SWE-Bench projects, no BRTs exist prior to a bug fix, and they are typically added as part of the pull request that introduces the fix. In the Defects4J dataset \cite{just2014defects4j}, only 4\% of the bug reports include a failing test case \cite{koyuncu2019ifixr}. Even in industrial settings, BRTs are usually deferred to the fix stage because bug reports often come from sources that lack the knowledge to create them at the time of reporting \cite{cheng2025agentic}. This implies that most bugs are reported without a reliable executable test, and developers are left with the task of reproducing the fault, which is time-consuming and challenging \cite{straubinger2023survey}, and could delay bug fixing. Automatically turning bug reports into tests can make it easier to debug, validate, and fix issues, as well as reduce the time developers spend reproducing failures. 

SWT-Bench \cite{mundler2024swt}, a benchmark for automated bug reproduction, demonstrates the growing potential of automatically generating BRTs. The current methods proposed for this task, which we discuss in more detail in the related work section (See Section \ref{sec:related_work}), have yet to achieve strong results. 
Directly prompting an LLM creates BRTs successfully in only 3.6\% of cases on SWT-Bench-Lite \cite{mundler2024swt}, a subset of SWT-Bench. Recent work shows that iterative prompting and multi-step interactions with LLMs can improve success rates on generation tasks \cite{nashid2025issue2test, pizzorno2024coverup, yang2024swe, xia2024automated}. However, in the context of bug reproduction, two challenges remain: determining whether (a) the test has no implementation problems or bugs in the test code itself \cite{zhu2024tddbench, ahmed2025otter, nashid2025issue2test, lin2024llms} (b) the test fails for the right reason and exercises the bug \cite{kang2023large, nashid2025issue2test, wang2024aegis}.
To address this, we introduce \textsc{AssertFlip}, a tool that generates bug-reproducing tests by first generating a test that passes on the buggy behaviour. If the test fails to run due to errors or setup issues, we refine it until it passes. Once we have a valid test that runs, we invert its logic to create a bug-revealing test. If the test cannot be fixed after a few rounds, we trigger a new regeneration attempt using the previous plan, test, and error to reflect on what went wrong in the earlier attempt.
% and then we invert it to make it fail when the bug isn't resolved. 
This pass-then-invert strategy can help avoid common failure modes in LLM-generated tests, such as broken syntax, setup errors, incomplete logic, or hallucinated assumptions.

\textit{Our key intuition is that LLMs are better at writing passing tests than at writing tests designed to crash or fail on purpose.} We show that \textsc{AssertFlip}\ outperforms prior methods at generating BRTs. Beyond these empirical gains, our findings point to a broader design paradigm for LLM-based bug reproduction. Prior work has treated failing-test generation as the default, but constraining the generation objective toward producing passing tests reshapes the entire workflow. Validation, coverage integration, and bug-report handling all need to be reconsidered under this premise. This shift defines a general design pattern of objective-driven generation, where explicit behavioral goals, such as “the test must pass” guide the structure of the workflow and open new directions for future research.

In summary, our main contributions of this paper are as follows:

\begin{enumerate}
    \item We propose a novel pass-then-invert test generation technique that helps the LLM focus on writing correct tests before transforming them into BRTs.
    \item We evaluate \textsc{AssertFlip}\ on SWT-Bench and show that it outperforms prior work on fail-to-pass success rate.
    \item We release our code and data to support reproducibility and to help others build on this work.
\end{enumerate}

\begin{figure*}[ht]
    \centering
    \includegraphics[width=\textwidth]{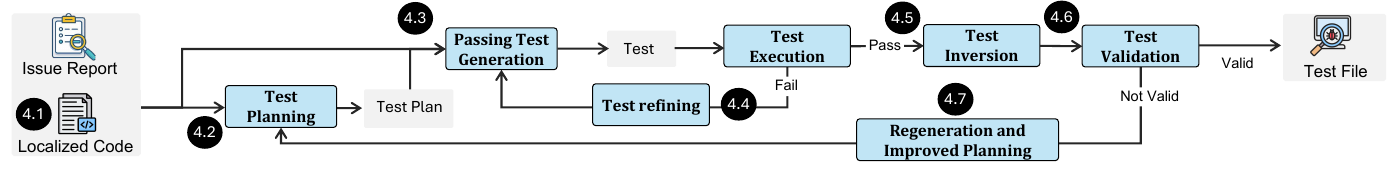}
    \caption{Overview of \textsc{AssertFlip}\ pipeline.}
    \label{fig:assertflip-pipeline}
\end{figure*}

\section{Related Work}\label{sec:related_work}

SWT-Bench \cite{mundler2024swt} is a recent benchmark developed specifically to evaluate the ability of LLMs to generate bug-reproducing tests. SWT-Bench has quickly become the standard benchmark for evaluating LLM-based bug reproduction, with many recent systems reporting results on it. It includes a public leaderboard on two subsets\footnote{\url{https://swtbench.com/?results=verified}}: SWT-Bench-Lite and SWT-Bench-Verified, the latter derived from SWE-Bench-Verified where human developers manually verified each instance to ensure a high-quality dataset \cite{openai2024swebench}. The benchmark was introduced alongside evaluations of several LLM-based code agents like SWE-Agent \cite{yang2024swe}, AutoCodeRover \cite{zhang2024autocoderover}, and Aider \cite{aider}, which were adapted with modified prompts for the task. The agents were evaluated on the fail-to-pass success rate, which is defined as the percentage of generated tests that fail on the buggy version of the code and pass after the corresponding fix is applied. This metric indicates whether a test correctly captures the buggy behaviour described in the issue and thus serves as a valid BRT. SWE-Agent+, a variant of SWE-Agent, achieved the highest fail-to-pass success rate reported in the SWT-Bench paper \cite{mundler2024swt}, with 19.2\% on SWT-Bench-Lite, compared to 15.9\% for SWE-Agent and just 3.6\% for direct LLM prompting. 

Amazon Q currently tops the leaderboard with 49\% on SWT-Verified and 37.7\% on SWT-Lite, though its architecture is undisclosed and does not rely on a single foundation model \footnote{\url{https://aws.amazon.com/q/}}. Since Amazon Q does not disclose any details and includes minimal traces, we refrain from directly comparing to it. However, we do report it for completeness. OpenHands \cite{wang2407openhands}, an open-source developer-style agent, achieves 27.7\% and 28.3\% on the verified and lite subsets, respectively, using Claude 3.5 Sonnet, despite not being specifically designed for bug reproduction.

A growing body of research has focused on LLM-driven methods for generating bug-reproducing tests from natural language issue descriptions. One of the earliest works in this area is LIBRO \cite{kang2023large}, which combines few-shot prompting, test post-processing, and heuristic ranking to generate BRTs. It was evaluated on the Defects4J and GHRB datasets \cite{lee2024github}. In the SWT-Bench paper \cite{mundler2024swt}, LIBRO was adapted for Python and evaluated on SWT-Bench-Lite, achieving 14.1\%. Otter \cite{ahmed2025otter} incorporates a self-reflective planner in which the LLM iteratively refines read/write/modify actions, then uses the final plan to guide test generation. Otter++ extends Otter by running multiple versions of the generation pipeline with different configurations and selecting the best test based on runtime feedback. This ensemble approach improves fail-to-pass rates on SWT-Verified from 31.4\% to 37.0\%, and on SWT-Lite from 25.4\% to 29.0\%. Issue2Test \cite{nashid2025issue2test} introduces a three-phase pipeline that first uses meta-prompting to extract project-specific test-writing guidelines, then performs root cause analysis, and generates test candidates. It enters an execution-feedback loop with two LLM components, one to classify test failures and another to verify that assertion failures match the original bug report. On SWT-Bench-Lite, it achieves a 30.4\% fail-to-pass rate. AEGIS \cite{wang2024aegis} introduces a two-agent framework for bug reproduction where a searcher agent retrieves relevant context and a reproducer agent generates and refines test scripts. Its key contribution is a finite-state machine controller that guides the reproducer through structured feedback loops, including syntax checks, execution results, and external verification. They report results only on the Lite subset, but we exclude them from our comparison in line with \textit{Otter} \cite{ahmed2025otter} due to unclear evaluation (they evaluate on \textit{SWE-Bench-Lite} but compare against results from \textit{SWT-Bench-Lite}, which is a different dataset).

Unlike the approaches mentioned above that attempt to generate failing tests and then attempt to analyze whether the failure is due to the intended bug or an unrelated issue, our approach starts by generating a passing test on the buggy version. If the test fails to run, we refine it until it executes successfully. We then invert its assertions to construct a BRT. \textsc{AssertFlip}\ achieves the best results amongst all known approaches on the leaderboard.

\section{Pass-first then invert}
Other work in bug reproduction prompts the LLM to write a failing test that exposes the bug, often augmenting this process with a self-reflective planner \cite{ahmed2025otter}, LLM-based validation \cite{nashid2025issue2test}, or execution and assertion matching loops \cite{kang2023large}. We hypothesize that LLMs perform better when writing correct tests than when asked to create a test that \emph{fails} on purpose \cite{mathews2024design}. We believe our method works primarily because we do not leave the responsibility of deciding when to stop and accept a test solely to the LLM. We adopt a more structured approach that allows us to better isolate the tasks of understanding the fault, creating a BRT, and validating the failure.

An LLM-generated test can fail for many reasons unrelated to the actual bug. These failures include a wide range of issues \cite{dou2024s, yuan2024evaluating}, such as syntax errors, import errors, and unintended top-level execution (e.g., code running before the test starts). Tests may also fail due to non-self-contained logic, missing setup or teardown steps, or uninitialized variables. The LLM may introduce outdated or hallucinated APIs/classes/functions, misuse testing frameworks, or generate placeholder code (e.g., assert False, \# TODO), or incomplete tests. Environment-level issues such as dependency mismatches, incorrect framework usage, or version mismatches can also cause a test to fail. If we rely solely on the LLM’s judgment to determine whether a test is valid and relevant to the bug, we risk accepting tests that fail for unrelated reasons, leading to false positives. While static analysis tools and linters such as \texttt{flake8} and \texttt{mypy} can help catch some of these issues, many deeper issues remain undetected. 

To mitigate this, we adopt a more controlled approach. Specifically, we require that any candidate test to pass on the buggy version of the program before it is evaluated further. This ensures that we avoid common sources of test failure that are due to artifacts of code generation rather than the bug itself. Thus, we filter out invalid tests before invoking the LLM’s reasoning to assess bug relevance. This approach allows us to retain some control when using LLMs. It is easier and more reliable to confirm that a test passes under buggy conditions than to reason about the potentially numerous causes of failure. In doing so, we narrow the list of things that can go wrong and focus on one task: determining if a syntactically valid executable that passes the test indeed exercises the behaviour associated with the bug. Once a valid passing test is obtained, we invert its assertions to construct a BRT. Further details of our tool are provided in the next section.

% \begin{figure*}[ht]
%     \centering
%     \includegraphics[width=\textwidth]{pipeline.pdf}
%     \caption{Overview of \textsc{AssertFlip}\ pipeline.}
%     \label{fig:assertflip-pipeline}
% \end{figure*}

% \begin{figure}[t]
% \centering
% \begin{tcolorbox}[mycustombox={Github Issue: astropy\_\_astropy-13236}]
% Consider removing auto-transform of structured column into \texttt{NdarrayMixin}. Currently, adding a structured \texttt{np.array} to a \texttt{Table} turns it into an \texttt{NdarrayMixin}:

% \begin{pycode}
% if (not isinstance(data, Column) and not data_is_mixin
%         and isinstance(data, np.ndarray) and len(data.dtype) > 1):
%     data = data.view(NdarrayMixin)
% \end{pycode}
  
% This was for serialization support. After \#12644, this may be unnecessary.  
% Proposal: emit \texttt{FutureWarning}, drop behaviour in v5.2.  
% (Issue summarized for brevity)
% \end{tcolorbox}
% \caption{Bug report for \texttt{astropy\_\_astropy-13236} from SWT-Bench.}
% \label{fig:astropy-bug}
% \end{figure}
\begin{figure}[t]
\centering
\includegraphics[width=\columnwidth]{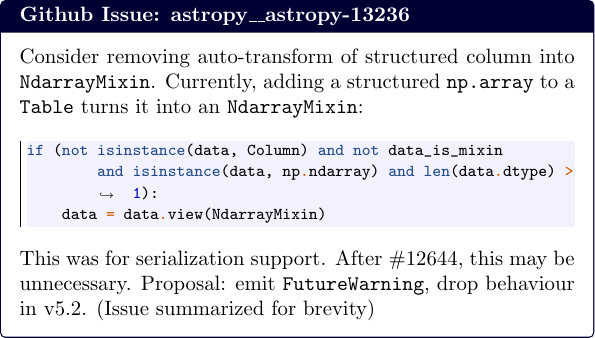}
\caption{Bug report for \texttt{astropy\_\_astropy-13236} from SWT-Bench.}
\label{fig:astropy-bug}
\end{figure}

\begin{figure}[t]
\centering
% \begin{tcolorbox}[mycustombox={Example Test Plan (Pass-First)}]
% \textbf{1. Test Setup:} Use \texttt{numpy} and the relevant parts of the \texttt{astropy} library.

% \vspace{0.4em}
% \textbf{2. Test Input:} Create a structured \texttt{np.array} like:
% \begin{pycode}
% np.array([(1, 2.0), (3, 4.0)], dtype=[('a', 'i4'), ('b', 'f4')])
% \end{pycode}

% \vspace{0.4em}
% \textbf{3. Bug Triggering:} Add the structured array to an \texttt{astropy.Table}. This causes a conversion to \texttt{NdarrayMixin}.

% \vspace{0.4em}
% \textbf{4. Test Structure:} Initialize the Table, add the array, and inspect the column type.

% \vspace{0.4em}
% \textbf{5. Assertions:} Check that the column type is \texttt{NdarrayMixin}. This is the incorrect behaviour but reflects the current bug.

% \vspace{0.4em}
% \textbf{6. Edge Cases:} Try variants with different field types or empty structured arrays.

% \vspace{0.4em}
% \textbf{7. Expected Outcome:} The test should pass for now, but will fail once the bug is fixed.
% \end{tcolorbox}
\includegraphics[width=\columnwidth]{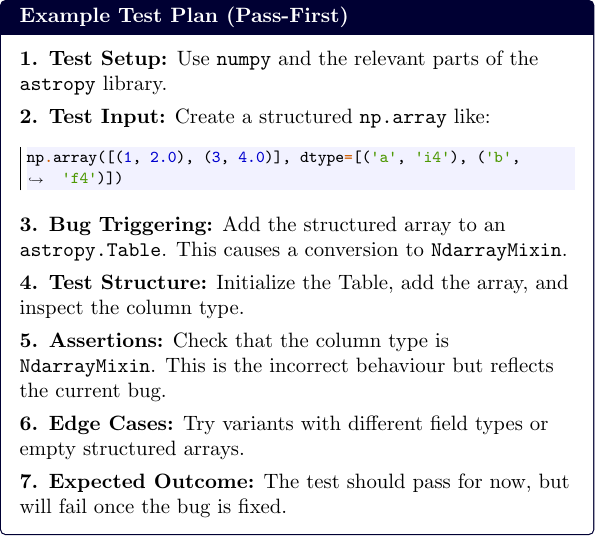}
\caption{Test plan generated for \texttt{astropy\_\_astropy-13236}.}
\label{fig:test-plan-astropy}
\end{figure}

\section{\textsc{AssertFlip}}\label{sec:approach}

This section describes \textsc{AssertFlip}, an LLM-based tool that generates BRTs from natural language bug reports and source code. 
To direct the LLM toward writing a valid executable test that exposes the bug, we ask it to write a test that passes on the buggy behaviour and provide it with the issue report alongside the localized buggy code. We continue refining the test using execution feedback until it passes. Rather than trying to handle all the possible reasons an LLM-generated test might fail, we first ensure the test passes. Once we have a correct working test code that reflects the bug, we invert it so the test now fails when the bug is present. Our intuition is that LLMs are better at writing passing tests than at writing failing ones, where the failure could happen for any number of unrelated reasons. This approach contrasts with prior methods proposed for bug reproduction like Otter, LIBRO, and Issue2Test, which attempt to produce failing tests directly. Figure \ref{fig:assertflip-pipeline} depicts the full pipeline. We describe each stage below.

\subsection{Localization input to the pipeline}

As a first step, existing tools typically begin by localizing buggy code. Among the current approaches proposed for this task, most rely on LLMs for localization. For example, Otter \cite{ahmed2025otter} uses a four-step process that first localizes test files and functions by prompting the LLM to pick the top-10 relevant test files based on the issue description, then lets the LLM pick relevant test functions within those test files, and repeats the same process for localizing focal files and focal functions. Issue2Test \cite{nashid2025issue2test} builds a hierarchical tree of the repository and provides it along with the issue report to the LLM to rank the most relevant files. AEGIS \cite{wang2024aegis} uses a Searcher Agent that employs an LLM to retrieve relevant code and test files based on the issue report using system commands and tool-specific interfaces. 

Since the focus of this paper is on creating tests from issues rather than bug localization, we decided to assess our pipeline using localization output from an existing tool called Agentless. \cite{xia2024agentless}. Agentless is a popular and, importantly, modular approach that we chose to obtain realistic localization data. The modular nature makes it possible to execute the localization phase alone, unlike most other tools in the space. For completeness, Agentless first identifies suspicious files by prompting a compact repository‐structure representation and refining this selection through embedding‐based retrieval. It then examines these files to isolate related code elements such as class declarations, function signatures, and variable definitions. Finally, it analyzes the actual code snippets of these elements to determine precise edit locations, which may be specified by line numbers, functions, or classes. Since Agentless generates multiple candidates at each stage, we adopt a conservative merging strategy that combines all proposed locations into a unified set.

\subsection{Test Planning}
Each run of \textsc{AssertFlip}\ begins with a planning phase, where we pass the relevant code snippets and issue description obtained from the previous step to the LLM to generate a detailed test plan before writing any code. The goal of this step is to help the model reason about how to reproduce the bug described in the issue before attempting to write the test itself. The LLM is instructed to plan a test that will \emph{pass} under the buggy version of the code, but still demonstrate the incorrect behaviour. Figure \ref{fig:test-plan-astropy} shows a test plan generated by the LLM for a bug in \texttt{astropy} and its corresponding GitHub issue (see Figure \ref{fig:astropy-bug}), where structured \texttt{np.array}s are incorrectly converted to \texttt{NdarrayMixin} when added to a Table. We selected this issue at random as a running example to demonstrate our approach.

\subsection{Passing Test Generation}

The core of our system is the pass-invert method. Instead of asking the LLM to write a failing test from scratch, which may lead to broken or non-executable tests, we first instruct the model to generate a \emph{passing} test that runs successfully but still exposes the bug. This test does not fail at runtime, but it reveals the incorrect behaviour described in the issue either through assertions that accept the buggy output or through comments explicitly identifying the flawed behaviour. Once the test is generated, we invert it by modifying its assertions such that it now fails when the bug is present and passes once the bug has been fixed. Using the plan from the previous step, we prompt the LLM to write a complete Python test using \texttt{pytest} or the testing framework used in the target project. We give it explicit constraints:
\begin{itemize}
    \item The test must pass when run against the buggy version.
    \item The test must clearly show the bug by asserting the buggy behaviour or use \texttt{pytest.raises(...)} or equivalent logic for exceptions
    \item The code should be executable with no placeholders or syntax errors.
\end{itemize}

\begin{figure}
\centering
% \begin{pycode}
% import pytest
% import numpy as np
% from astropy.table import Table
% from astropy.table.ndarray_mixin import NdarrayMixin

% def test_structured_array_conversion_to_ndarraymixin():
%     structured_array = np.array(
%         [(1, 2.0), (3, 4.0)],
%         dtype=[('a', 'i4'), ('b', 'f4')]
%     )
%     table = Table([structured_array])
%     column_type = type(table.columns[0])
%     # BUG: this is incorrect behaviour but currently happens
%     assert column_type is NdarrayMixin
% \end{pycode}
\includegraphics[width=\columnwidth]{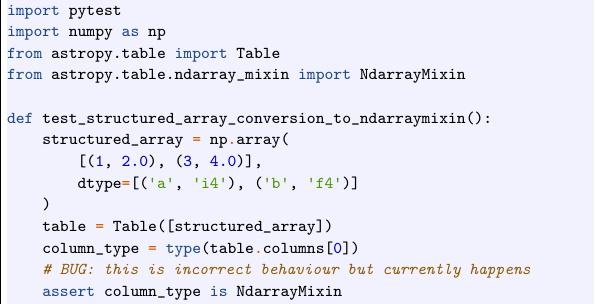}
\caption{The passing test generated by the LLM reflecting the buggy behavior.}
\label{fig:llm-passing-test}
\end{figure}

Figure \ref{fig:llm-passing-test} shows the passing test that the LLM generated based on the aforementioned plan in Figure \ref{fig:test-plan-astropy}. This test passes, but only because the bug is still present, the conversion to \texttt{NdarrayMixin} is happening, and the test explicitly checks for that.

\subsection{Test Refinement Loop}
\label{sec:refinement}

\begin{figure}
\centering
% \begin{tcolorbox}[mycustombox={Error Prompt}]
% Executing the test yields the error shown below. Modify or rewrite the test to correct it.

% \vspace{0.5em}
% \textbf{Test Code:}
% \begin{pycode}
% import pytest
% import numpy as np
% from astropy.table import Table
% from astropy.table.column import NdarrayMixin

% def test_structured_array_conversion_to_ndarraymixin():
%     structured_array = np.array(
%         [(1, 2.0), (3, 4.0)],
%         dtype=[('a', 'i4'), ('b', 'f4')]
%     )
%     table = Table([structured_array])
%     column_type = type(table.columns[0])
%     # BUG: this is incorrect behaviour but currently happens
%     assert column_type is NdarrayMixin
% \end{pycode}

% \vspace{0.4em}
% \textbf{Error Message:}  
% \begin{textformat}
% ImportError: cannot import name 'NdarrayMixin' from 'astropy.table.column'
% \end{textformat}
% \end{tcolorbox}
\includegraphics[width=\columnwidth]{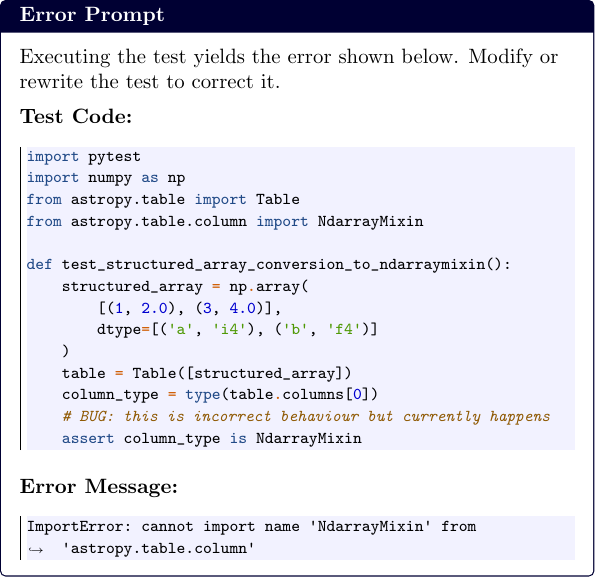}
\caption{Error Prompt for \texttt{astropy\_\_astropy-13236}}
\label{fig:error-prompt}
\end{figure}

Generated tests often fail to execute correctly on the first attempt. This could be due to a variety of reasons: missing imports, incorrect assumptions, setup errors, or subtle misunderstandings of the code. Rather than discarding these tests, \textsc{AssertFlip}\ enters a \textit{refinement loop}. In this loop, we prompt the LLM to revise the test based on the precise \textit{error message} that occurred during execution. As long as the test is making progress (e.g., different errors are appearing), we continue refining it, up to a maximum of ten iterations. If a valid passing test is still not produced within this limit, the process moves to the regeneration and improved planning phase described in Section~\ref{sec:regeneration}. This iterative strategy often allows the LLM to fix import paths, correct argument usage, or add missing test setup, without losing the structure of the originally generated test. An example of the prompt used during refinement is shown in Figure~\ref{fig:error-prompt}.

\subsection{Inverting the Passing Test}

\begin{figure}
\centering
% \begin{pycode}
% import pytest
% import numpy as np
% from astropy.table import Table
% from astropy.table.column import Column

% def test_structured_array_conversion_to_column():
%     structured_array = np.array(
%         [(1, 2.0), (3, 4.0)],
%         dtype=[('a', 'i4'), ('b', 'f4')]
%     )
%     table = Table([structured_array])
%     column_type = type(table.columns[0])
%     # This is the expected behaviour after the fix
%     assert column_type is Column
% \end{pycode}
\includegraphics[width=\columnwidth]{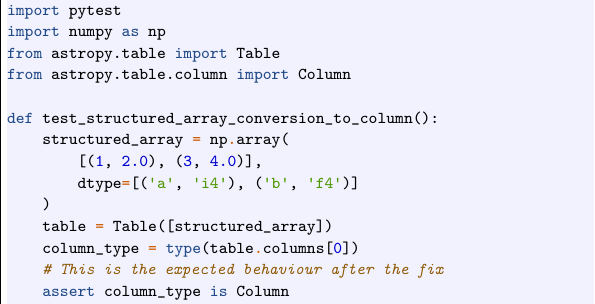}
\caption{The inverted test now fails when the bug is present.}
\label{fig:llm-inverted-test}
\end{figure}

Once we have generated a passing test that compiles, runs, and exercises code relevant to the bug, we flip its logic. This test already demonstrates the bug, as it passes because the system exhibits the buggy behaviour. At this point, we do not need to rewrite or redesign the test. Instead, we keep the structure the same and update the assertions to reflect the correct behaviour. We pass the issue report and prompt the LLM to do this with very specific instructions: change as little as possible, keep the test readable and minimal, and modify the logic so that the test fails when the bug is present and passes only when the bug is fixed. Any comments that previously acknowledged the buggy behaviour (e.g., \# BUG: this is incorrect) are removed. Because the test is already valid and it compiles, executes cleanly, and covers the suspect lines, inversion becomes a simple step of changing assertions or outputs. Figure \ref{fig:llm-inverted-test} shows the inverted version of the test for \texttt{astropy\_\_astropy-13236}. This test now serves as a proper BRT: it fails on the buggy version of the code and will pass after the fix is applied.

\subsection{Test Validation}

\begin{figure}
\centering 
% \begin{tcolorbox}[mycustombox={Validation Prompt: astropy\_\_astropy-13236}]
% \textbf{Task:}  
%  Validate whether the test correctly reproduces the bug and whether it is a valid test case. If the failure is not due to the bug, you must provide a detailed explanation of why the test is incorrect and doesn't reproduce the bug. 

% \textbf{You are given:}
% an issue report,  a test that fails, and the corresponding error message.

% \vspace{0.5em}
% \textbf{GitHub Issue:}  
% Adding a structured \texttt{np.array} to an \texttt{astropy.Table} wraps it in an \texttt{NdarrayMixin}. This behaviour is no longer needed and will be deprecated in v5.2. (summarized)

% \vspace{0.4em}
% \textbf{Test Execution Output:}
% \vspace{0.3em}

% \begin{textformat}
% AssertionError: 
% assert <class 'astropy.table.ndarray_mixin.NdarrayMixin'> is Column
% \end{textformat}

% \vspace{0.4em}
% \textbf{Expected Response:}
% \begin{jsonformat}
% {
%   "revealing": true,
%   "reason": "The test exposes the current behaviour where structured arrays are cast to NdarrayMixin. This matches the behaviour described in the issue."
% }
% \end{jsonformat}
% \end{tcolorbox}
\includegraphics[width=\columnwidth]{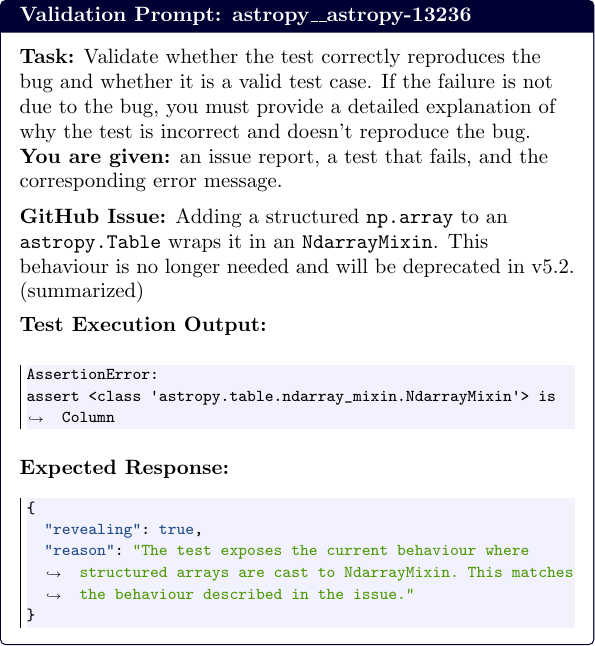}
\caption{Prompt used to validate whether a test correctly reveals the bug.}
\label{fig:validation-prompt}
\end{figure}

To ensure the test reveals the reported bug, we perform an LLM-based validation step. We run the test against the buggy version of the code, and pass the observed error trace along with the issue description into a separate LLM validation prompt shown in Figure \ref{fig:validation-prompt}. The validator is asked to determine whether the failure is caused by the bug described in the report or due to unrelated issues. If validation passes, the test is accepted. If the test fails validation, we trigger a new generation cycle, this time asking the LLM to rethink its plan entirely and try a new strategy for exposing the bug.

\subsection{Regeneration and Improved Planning} \label{sec:regeneration}

If the system repeatedly generates tests that either fail or are rejected during validation, we do not allow the LLM to simply tweak the same test or revise its last response. Instead, we trigger a full \textbf{regeneration} phase. In this mode, the LLM is instructed to abandon its previous strategy and adopt a different approach. The idea of this design is that if the LLM’s initial reasoning is flawed, iterative refinements based on that flawed strategy are unlikely to yield valid results. To help it learn from its earlier mistakes, we include:
\begin{itemize}
  \item The original bug report \texttt{<ISSUE TICKET>} and code snippets \texttt{<CODE SNIPPETS>}
  \item The previous plan \texttt{<THOUGHT PROCESS>}
  \item The failed test attempt \texttt{<TEST ATTEMPT>} and the error message \texttt{<ERROR>}
  \item The feedback on why the test was rejected \texttt{<FEEDBACK>}
\end{itemize}

\begin{figure}
\centering 
% \begin{promptbox}
% \textbf{Instructions:}

% You are an expert senior Python test-driven developer tasked with assisting your junior who is unable to write tests that reveal reported bugs. Your goal is to plan the creation of test functions that PASS but still expose the reported bug. 

% You will be provided with an ISSUE TICKET and a set of CODE SNIPPETS which might contain the buggy logic. You will also be given the THOUGHT PROCESS of your junior who was trying to write the test, the TEST ATTEMPT they wrote, the ERROR it produced. As well as FEEDBACK explaining why the previous test failed.

% Your task is to analyze the described problem and previous attempt in detail and create a new PLAN for writing the test.

% You MUST NOT reuse or copy the previous plan. You may use it to understand what failed, but you must take a different angle that avoids the same mistakes.

% \end{promptbox}
\includegraphics[width=\columnwidth]{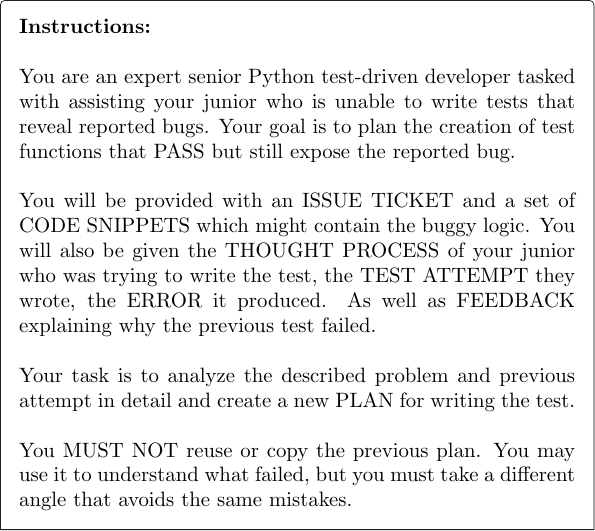}
\caption{Instructions given in regeneration prompt.}
\label{fig:reg-prompt}
\end{figure}

The feedback is generated either during the validation step if the test does not correctly expose the bug or when the system fails to produce a passing test despite multiple refinement attempts, whether by repeatedly triggering the same error or exhausting the retry limit. The model is then prompted to rethink its strategy from scratch. We explicitly instruct the model not to reuse its earlier plan or structure. The prompt, shown in Figure \ref{fig:reg-prompt}, instructs the LLM to reflect on what went wrong in the previous attempt and generate a completely new plan that avoids the previous pitfalls \cite{gou2305critic}. This gives the system a second chance and encourages the LLM to diversify its reasoning and explore alternative test designs that might better expose the bug. 

% In our implementation, this “re-planning” prompt assigns the LLM a persona \cite{white2023prompt} as a senior developer mentoring a junior developer whose previous test plan failed, it must now propose a different strategy. 

\subsection{Additional Utilities: Code Retrieval via \texttt{get\_info}} \label{sec:getinfo}

\begin{figure}
\centering
% \begin{tcolorbox}[mycustombox={Example Usage of \texttt{get\_info}}]

% \textbf{Tool Call:}
% \begin{jsonformat}
% {
%   "name": "get_info",
%   "path": "astropy/table/table.py",
%   "name": "Table._convert_data_to_col",
%   "line": 1179
% }
% \end{jsonformat}

% \vspace{0.5em}
% \textbf{Expected Output:}
% \begin{pycode}  
% class Table:
%     ...

%     def _convert_data_to_col(self, data, copy=True, default_name=None, dtype=None, name=None):
%         """
%         Convert any allowed sequence data ``col`` to a column 
%         ... (omitted for brevity)
%         """
%         data_is_mixin = self._is_mixin_for_table(data)
%         ... (omitted for brevity)
%         return col
% \end{pycode}
% \end{tcolorbox}
\includegraphics[width=\columnwidth]{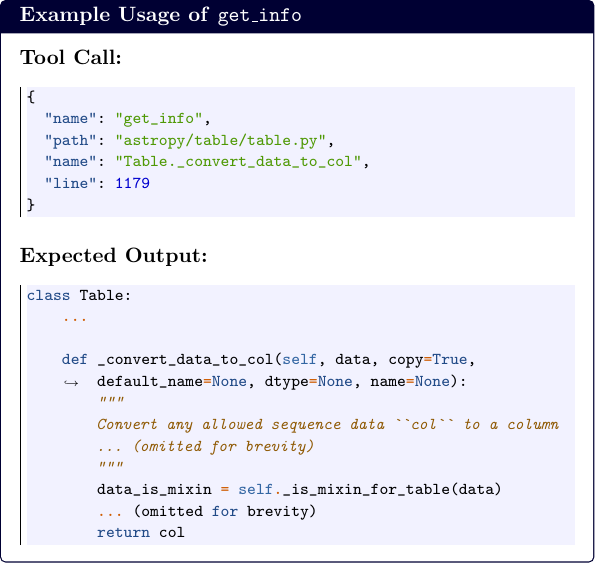}
\caption{An example call to \texttt{get\_info} for retrieving context.}
\label{fig:get_info_example}
\end{figure}

LLMs often struggle when working with partial code snippets, frequently hallucinating and making incorrect assumptions about missing code. This leads to errors during tasks like test generation or bug fixing. To reduce these errors, we introduce a tool function that allows the LLM to request additional information about any names in the excerpt, such as functions, classes, or variables. The tool is implemented using OpenAI’s function-calling interface \cite{openai-function-calling}. At any point in the conversation, the model can request additional information about a symbol (such as code artifacts like functions or classes). The tool then performs static analysis to locate the symbol’s definition and returns a trimmed, valid Python excerpt before continuing the conversation. To keep responses compact, less relevant sections are omitted and replaced with ellipses (\texttt{...}). The tool follows import paths and inheritance chains when needed and can merge context from multiple modules. This is especially useful in scenarios like refining a failing test. If the test references a function or variable with unclear behaviour, the tool can be used to provide enough context for generating a correct fix. Figure \ref{fig:get_info_example} shows an example of how the \texttt{get\_info} tool is used to retrieve the definition of a method. This tool is available in the planning phases, test generation, and error fixing.

\section{Evaluation}\label{sec:evaluation}

\subsection{Experimental Setup}
\subsubsection{\textbf{Benchmark}}
We use SWT-Bench \cite{mundler2024swt} for our evaluation and run our tool on two datasets: SWT-Bench-Lite and SWT-Bench-Verified. Both are derived from real-world issue reports and patches in 12 popular open-source Python projects on GitHub. Each instance includes a natural language bug report, a corresponding fix patch, and a test that fails on the buggy version and passes after the fix is applied. The two datasets differ primarily in size and the strictness of their selection criteria. SWT-Bench-Lite contains 276 instances where only a single file is edited in the fix. SWT-Bench-Verified includes 433 instances that human developers have manually validated to ensure that each issue is clearly stated, the patch is meaningful, and the test accurately reflects the fix, making it more reliable than the other subsets and thus our choice for focusing our evaluation on. We also include SWT-Bench-Lite in our evaluation because it is commonly used in prior work, which allows for easier comparison against the existing methods.

\subsubsection{\textbf{LLM}} We use OpenAI's GPT-4o (gpt-4o-202408-06) for our experiments so that we can fairly compare against the other submissions on SWT-Bench, which utilize models from the GPT-4 family that have cutoffs before October 2023, when SWE-Bench (the underlying dataset behind SWT-Bench) was released.

\subsubsection{\textbf{Evaluation Metrics}}

To assess the effectiveness of our test generation approach, we adopt the evaluation metrics introduced by the SWT-Bench paper \cite{mundler2024swt}. Specifically, we use two metrics:
\begin{itemize}
    \item \textbf{Fail-to-Pass (F$\rightarrow$P) Success Rate}: This metric measures the proportion of instances where at least one generated test fails on the buggy version but passes after the corresponding golden patch is applied. Such F$\rightarrow$P tests are also used in prior work to indicate successful reproduction of issues and are also important in validating bug-fix correctness. A successful instance must contain at least one F$\rightarrow$P test and no tests that fail after the patch ($\times\rightarrow$F).
    
    \item \textbf{Delta Mean Change Coverage ($\Delta C$)}: This metric specifically measures how well the generated tests cover the lines of code that were modified by the golden patch. It is computed as the percentage of modified (added, removed, or edited) lines that are newly covered by the generated tests. 
\end{itemize}

\subsubsection{\textbf{Baselines}}
We evaluate \textsc{AssertFlip}\ against systems reported on the SWT-Bench leaderboard for the Verified subset.\footnote{\url{https://swtbench.com/?results=verified}} This includes two baselines introduced by Mündler et al. \cite{mundler2024swt}: \textit{ZeroShotPlus}, which uses direct LLM prompting and generates a novel code diff format introduced by their paper, and \textit{LIBRO} \cite{kang2023large}, an earlier test generation system re-evaluated on their dataset and uses the same proposed patch. \textit{Otter} \cite{ahmed2025otter} a recent approach for automated bug reproduction, and its variant \textit{Otter++}, which selects outputs from five different prompting methods. Also listed is \textit{OpenHands} \cite{wang2407openhands}, an open-source agent that was adapted for bug reproduction, and \textit{Amazon Q} \cite{amazonqweb}, a commercial system with results reported on the leaderboard, though its underlying setup is undisclosed. OpenHands uses Claude 3.5 Sonnet, and Amazon Q uses multiple foundation models, making their results not directly comparable to the other systems.

As for the Lite subset, we include results from five methods introduced in the SWT-Bench paper \cite{mundler2024swt}: \textit{ZeroShotPlus}, \textit{LIBRO}, \textit{AutoCodeRover}, \textit{SWE-Agent}, and its variant \textit{SWE-Agent+}. We also compare to prior works in the literature, such as \textit{Otter}, its variant \textit{Otter++}, and \textit{Issue2Test}. Additional systems listed on the public leaderboard include \textit{OpenHands} and \textit{Amazon Q}.

\subsection{Effectiveness of \textsc{AssertFlip}\ against the Baselines}

Table~\ref{tab:swebench-verified} reports the F$\rightarrow$P rates and $\Delta$ Coverage scores for  \textsc{AssertFlip}\ and prior test generation systems on SWT-Bench-Verified.  \textsc{AssertFlip}\ achieves a 43.6\% F$\rightarrow$P success rate, successfully resolving 189 F$\rightarrow$P tests out of 433 issues and outperforming all comparable baselines. Alongside its higher F$\rightarrow$P rate,  \textsc{AssertFlip}\ also achieves higher $\Delta$ Coverage score, indicating that the generated tests exercise a larger portion of the buggy code compared to those from other systems. Following SWT-Bench, we define an instance as \textbf{resolved} if the \textit{generated} test fails on the buggy version and passes on the fixed version. Otherwise, it is \textbf{unresolved}. We adopt this terminology throughout the evaluation.

While our reported score reflects performance over all 433 Verified issues, it is important to note that, unlike other systems, our approach deliberately abstains from generating solutions for every instance. We only generate 326 of the 433 cases. This is not a limitation but a feature, as our tool avoids producing low-confidence or misleading outputs, resulting in fewer incorrect responses that could waste developer time. This aligns with the motivation behind BouncerBench by Mathews et al. \cite{mathews2025your}, which highlights the importance of abstention in automated software engineering systems, arguing that "sometimes no answer is better than a wrong one." We believe this enhances the trustworthiness of our system. As a result, our actual resolution rate is 58\% (189 out of 326). 

Among the baselines, ZeroShotPlus performs the lowest, resolving only 62 bugs. LIBRO shows a modest improvement at 17.8\%, resolving 15 more bugs than direct prompting. Otter outperforms these techniques at 31.4\% F$\rightarrow$P rate, which increases to 37.0\% with Otter++. The improvement comes from running the test generation stage five times with different heterogeneous prompts. Our approach,  \textsc{AssertFlip}, which generates passing tests and then inverts them, outperforms all these methods. This result supports the hypothesis that generating valid passing tests and flipping their oracle is more reliable than attempting to generate failing tests directly. Compared to Amazon Q, the current top-ranked system on the leaderboard,  \textsc{AssertFlip}\ performs closely, despite using only a single model. Amazon Q achieves a 49.0\% F$\rightarrow$P rate, representing a five percentage point advantage in success rate. However, direct comparison is limited since we know nothing about how Amazon's Q developer works.

Table~\ref{tab:swebench-lite} reports the results on SWT-Bench-Lite.  \textsc{AssertFlip}\ achieves a 36\% F$\rightarrow$P success rate, outperforming all other open and publicly described approaches. This includes Otter++, AutoCodeRover, and SWE-Agent+, which all use different architectures. Our method falls just one percentage point behind Amazon Q, the top-performing system on the public leaderboard. This result highlights the competitiveness of our pass-then-invert strategy.

\begin{table}[t]
\centering
\small
% \footnotesize
\begin{tabular}{lccc}
\toprule
\multicolumn{1}{c}{\textbf{Approach}} & 
\multicolumn{2}{c}{\textbf{F$\rightarrow$P}} & 
\multicolumn{1}{c}{\textbf{$\Delta$ Coverage (\%)}} \\
\cmidrule(lr){2-3}
 & \textbf{Total} & \textbf{Rate} & \\
\midrule
ZeroShotPlus (GPT-4o)          & 62   & 14.3 & 34.0 \\
LIBRO (GPT-4o)                 & 77   & 17.8 & 38.0 \\
OpenHands\textsuperscript{*} (Claude 3.5 Sonnet)              & 120  & 27.7 & 52.9 \\
Otter (GPT-4o)               & 136  & 31.4 & 37.6 \\
Otter++ (GPT-4o)                & 160  & 37.0 & 42.8 \\
\rowcolor{gray!10}
\textbf{\textsc{AssertFlip}} (GPT-4o)   & 189  & 43.6 & 49.1 \\
Amazon Q (Amazon Bedrock) \textsuperscript{*} & 212 & 49.0 & 57.4 \\
\bottomrule
\end{tabular}
\vspace{0.6em}
\caption{Comparison with prior methods on SWT-Bench-Verified. \textsuperscript{*}Results not obtained under comparable settings.}
\label{tab:swebench-verified}
\end{table}

\begin{table}[t]
\centering
% \small
\begin{tabular}{lccc}
\toprule
\multicolumn{1}{c}{\textbf{Approach}} & 
\multicolumn{2}{c}{\textbf{F$\rightarrow$P}} \\
\cmidrule(lr){2-3}
 & \textbf{Total} & \textbf{Rate} & \\
\midrule
AutoCodeRover (GPT-4)         & 25   & 9.1  \\
ZeroShotPlus (GPT-4)         & 28   & 10.1  \\
LIBRO (GPT-4)                 & 42   & 15.2  \\
SWE-Agent (GPT-4)              & 46   & 16.7  \\
SWE-Agent+ (GPT-4)             & 53   & 19.2  \\
Otter (GPT-4o)                 & 70  & 25.4  \\
OpenHands (Claude 3.5 Sonnet) & 78  & 28.3  \\
Otter++ (GPT-4o)               & 80  & 28.9 \\
Issue2Test (GPT-4o-mini)   & 84 & 30.4 \\
\rowcolor{gray!10}
\textbf{\textsc{AssertFlip}} (GPT-4o)   & 99  & 36 \\
Amazon Q (Amazon Bedrock) & 104 & 37.7  \\
\bottomrule
\end{tabular}
\vspace{0.6em}
\caption{Comparison with prior methods on SWT-Bench-Lite.}
\label{tab:swebench-lite}
\end{table}

\subsection{Ablation Study}

To understand the contribution of individual components in our tool, we conduct a series of ablation experiments. Tables ~\ref{tab:ablation-assertflip} report F$\rightarrow$P success rates under different configurations of the pipeline, including removal of the LLM-based validator, omission of the planning step, and the use of perfect localization. We also include a variant that retains the full pipeline but prompts the LLM to generate failing tests directly instead of following the pass-then-invert strategy. In this direct-fail variant, the validation stage is moved after test generation and execution to enable iterative refinement and ensure a fair comparison with the original pipeline. Table~\ref{tab:ablation-regens} shows variation in the number of regenerations. 

The direct-fail variant represents one of the most important ablations demonstrating the effectiveness of our approach. Although it uses the full pipeline, prompting the LLM to generate failing tests directly rather than passing tests that are then inverted, results in a substantial drop in performance, resolving only 105 instances compared to 189 in the pass-then-invert configuration. This corresponds to a decline of over 19 percentage points in F$\rightarrow$P success rate, supporting our core intuition that LLMs are more reliable when asked to generate passing tests rather than failing tests directly.

Removing LLM validation results in a slight decrease in F$\rightarrow$P success rate from 43.6\% to 38.5\%. Although there is a drop, we observe that the results are still superior to those of all comparable methods. While the number of generated tests increases from 326 to 352, the number of successful instances drops (189 $\rightarrow$ 167), and the number of unresolved cases increases (137 $\rightarrow$ 185). This suggests that the validation step plays a role in filtering and rejecting false positives that would otherwise be accepted. Moreover, when a test fails validation, it triggers a new regeneration cycle that often lead to successful test generation. Thus, the validation step is essential in filtering poor tests through iterative regeneration.

When removing the planner and re-planning phase, we observe an increase in the number of tests generated by our tool to 351. However, the number of successfully resolved instances drops to 181. This suggests that LLMs benefit from following a structured plan, which may help them stay on track and produce tests that are more likely to succeed. Although the drop in performance is relatively small, the overall accuracy of our tool, measured as the number of resolved instances out of the total generated, decreases from 58\% to 51.6\%. Since we care about reducing the generation of invalid tests, we retain the planner as it helps guide the LLM. Despite this, our system still outperforms all prior GPT-4o systems, further demonstrating the strength of our pass-then-invert technique.

We also evaluate the impact of localization quality by comparing performance under realistic (Agentless) and perfect localization. In our perfect localization setup, we use the Git patch from the fix in the SWT-Bench dataset \cite{mundler2024swt} to localize the buggy code at both the file and line levels. To balance conciseness and completeness, we adopt a skeleton format: we treat each source file as an abstract syntax tree (AST) and collapse non-essential nodes, such that the edited lines are presented alongside their enclosing structures (classes, functions, blocks). This representation remains compact yet preserves the surrounding context.  We observe that the total number of generated instances remains constant across localization settings, although the set of bugs that get resolved changes. Perfect localization resolves 24 instances that are unresolved under the realistic localization configuration. On the other hand, realistic localization successfully resolves 23 instances that remain unresolved under perfect localization. We also found 11 cases where tests were generated and resolved under perfect localization, but no tests were generated when using the realistic localization setup. These differences highlight how test generation is highly sensitive to localization quality and that different localization methods expose different sets of bugs. Importantly, even with realistic localization, which is much closer to what we would expect in real-world settings, our tool maintains strong performance and resolves a large number of instances. This suggests that the effectiveness of  \textsc{AssertFlip}\ does not depend on precise localization.

Table~\ref{tab:ablation-regens} shows how the number of regeneration attempts affects performance. Allowing more regenerations leads to an increase in the number of bugs resolved. Without any regenerations, the system solves 141 instances. This increases to 169 with five regenerations and 189 with 10, highlighting the value of re-planning as it gives the model multiple chances to expose the bug. 

\begin{table}[t]
\centering
\small
% \footnotesize
\begin{tabular}{lccc}
\toprule
\textbf{Variant} & \textbf{Generated Tests} & \textbf{ F$\rightarrow$P} & \textbf{Rate (\%)} \\
\midrule
\rowcolor{gray!10}
\textbf{\textsc{AssertFlip}}       & 326 & 189 & 43.6 \\
Without LLM Validation                & 352 & 167 & 38.5 \\
Without Planner                       & 351 & 181 & 41.8 \\
With Perfect Localization          & 326 & 189 & 43.6 \\
\midrule
Direct-Fail Variant  & 330 & 105 & 24.2 \\
\bottomrule
\end{tabular}
\vspace{0.5em}
\caption{Ablation study showing generation and F$\rightarrow$P performance across system components.}
\label{tab:ablation-assertflip}
\end{table}

\begin{table}
\centering
\small
% \footnotesize
\begin{tabular}{lccc}
\toprule
\textbf{Regenerations} & \textbf{Generated Tests} & \textbf{F$\rightarrow$P Total} & \textbf{F$\rightarrow$P Rate (\%)} \\
\midrule
0   & 219 & 141 & 32.5 \\
5   & 300 & 169 & 39.0 \\
10\textsuperscript{†} & 326 & 189 & 43.6 \\
\bottomrule
\end{tabular}
\vspace{0.5em}
\caption{Impact of regeneration attempts on generation and F$\rightarrow$P success. \textsuperscript{†}10 regenerations is the default configuration.}
\label{tab:ablation-regens}
\end{table}

\subsection{Cost Effectiveness of \textsc{AssertFlip}}

The average cost of running \textsc{AssertFlip}\ per instance on SWT-Bench-Verified depends primarily on the number of regeneration attempts. At zero regenerations, the average cost is approximately 18 cents (0.1812 USD) per instance. This increases to 60 cents (0.6018 USD) with five regeneration attempts, and reaches 1.006 USD at ten regenerations. This includes all LLM interactions required throughout the entire pipeline: planning, test generation, test refinement, inversion, and validation. All computations use OpenAI’s GPT-4o pricing at the time of evaluation. The cost at the 10-regeneration setting, which is our default, remains similar to those reported for other unit test generation systems, bug reproduction tools, and LLM-based code agents. For users with cost constraints, running the system with five regeneration attempts offers a strong balance. The performance remains high and still outperforms all known tools on SWT-Bench-Verified, while cutting the cost by roughly 40\% compared to the 10-regeneration setting. During generation, \textsc{AssertFlip}\ accumulates context within each regeneration attempt to guide iterative improvements. Still, this context is discarded when a new regeneration cycle begins, and only the previous plan, test, and error message are passed. This design helps control prompt size and cost while preserving reasoning during each attempt. 

We use the cost figures reported in the SWT-Bench's paper \cite{mundler2024swt}, and compare them on the SWT-Bench-Lite subset, which consists of 276 instances. At 10 regeneration attempts, running \textsc{AssertFlip}\ across all 276 instances would cost approximately \$266.7 and an average cost of \$0.96 per instance. In contrast, ZeroShot and ZeroShotPlus cost around \$82, while LIBRO \cite{kang2023large} costs \$420. SWE-Agent \cite{yang2024swe} and SWE-Agent+ are reported at \$290.71 and \$478.21, respectively. AutoCodeRover \cite{zhang2024autocoderover} is reported at \$368.4. These comparisons show that even at its highest regeneration setting, \textsc{AssertFlip}\ remains similar to or lower than most other approaches while delivering stronger performance.

To better understand where costs are incurred, we analyzed the total and per-instance cost of running \text{\textsc{AssertFlip}} on all 433 verified instances from SWE-bench-Verified. The overall cost was \$435.92, averaging \$1.00 per instance. However, this average masks substantial variance across projects. For example, \texttt{django} dominated the cost profile, accounting for over 70\% of the total (\$309.02), with an average of \$1.43 per instance across 216 instances. In contrast, projects like \texttt{sympy} and \texttt{scikit-learn} had much lower average costs per instance (as low as \$0.13). A detailed cost breakdown per project is provided in Table~\ref{tab:assertflip-cost-breakdown}. This indicates that costs vary across projects and may depend on project-specific characteristics. To understand the source of cost variation, we initially examined factors such as fix difficulty, bug report length, and localization context, but found no clear correlation. A closer analysis of the execution traces later revealed that projects like \texttt{django} and \texttt{sphinx}, which showed the highest per-instance costs, have more project-specific mistakes and therefore more regenerations. We find that these two projects have more custom setups, like specific testing setup requirements, which LLMs might not be familiar with, leading to additional regenerations and higher overall cost.

\begin{table}[ht]
\centering
% \small
\footnotesize
\caption{Per-project cost analysis from running \textsc{AssertFlip}\ on 433 verified instances from SWT-Bench-Verified.}
\begin{tabular}{lrrr}
\toprule
\textbf{Project} & \textbf{Instances} & \textbf{Avg. Cost (USD)} & \textbf{Total Cost (USD)} \\
\midrule
\texttt{django}         & 216 & 1.43  & 309.02 \\
\texttt{sphinx}                  & 28  & 1.84  & 51.63  \\
\texttt{sympy}          & 73  & 0.37  & 26.73  \\
\texttt{matplotlib} & 32  & 0.42  & 13.29  \\
\texttt{pytest}                  & 15  & 0.84  & 12.64  \\
\texttt{astropy}      & 17  & 0.70  & 11.95  \\
\texttt{pylint}                  & 6   & 0.61  & 3.65   \\
\texttt{scikit-learn}                  & 24  & 0.13  & 3.17   \\
\texttt{xarray}        & 15  & 0.16  & 2.47   \\
\texttt{requests}         & 4   & 0.23  & 0.92   \\
\texttt{seaborn}      & 2   & 0.20  & 0.39   \\
\texttt{flask}        & 1   & 0.07  & 0.07   \\
\midrule
\textbf{Total / Average}         & \textbf{433} & \textbf{1.00} & \textbf{435.92} \\
\bottomrule
\end{tabular}
\label{tab:assertflip-cost-breakdown}
\end{table}

\section{Discussion}

\subsection{Does fail-to-pass rate tell the whole story?}

\begin{figure}[t]
    \centering
    \includegraphics[width=\columnwidth]{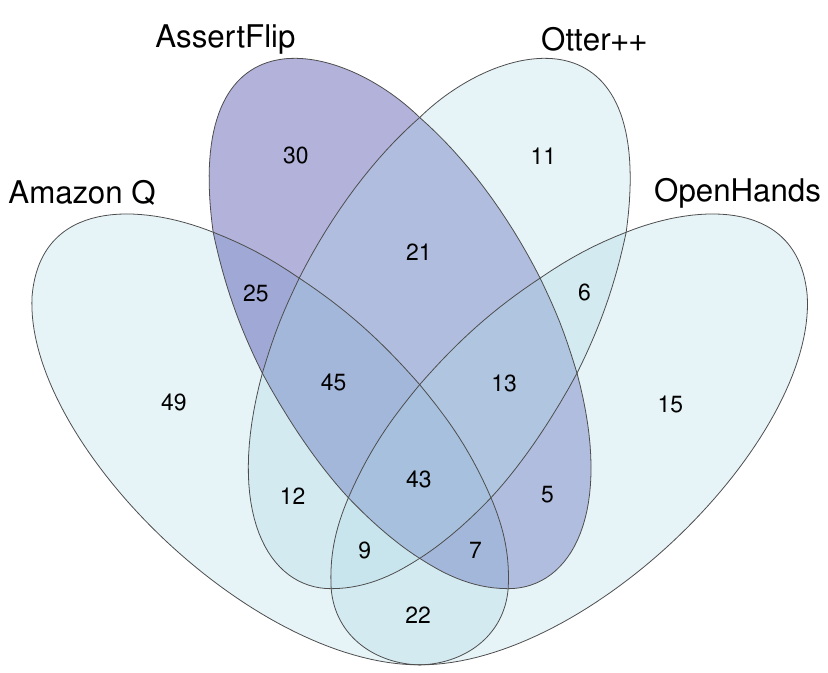}
    \caption{Overlap of resolved bugs across \textsc{AssertFlip}\, Amazon Q, Otter++, and OpenHands on SWT-Bench-Verified.}
    \label{fig:venn_overlap}
\end{figure}

While the F$\rightarrow$P rate is commonly used as the primary metric in bug reproduction benchmarks, it does not fully capture the distinct capabilities of different systems. In Figure~\ref{fig:venn_overlap}, we visualize the overlap of resolved instances among \textsc{AssertFlip}, Amazon Q, Otter++, and OpenHands on the SWT-Bench-Verified subset. \textsc{AssertFlip}\ resolves 30 bugs that none of the other three systems handle. On the other hand, the other baselines each resolve instances that \textsc{AssertFlip}\ misses. Amazon Q resolves 49 unique bugs, Otter++ resolves 11, and OpenHands resolves 15, with additional overlaps between them. The four systems collectively resolve 313 bugs, corresponding to a F$\rightarrow$P rate of 72.2\%, significantly surpassing the highest reported single-system F$\rightarrow$P rate of 45\% for Amazon Q.

This pattern indicates that these systems are often solving different types of issues, and that F$\rightarrow$P rate alone can hide that. These differences may reflect variations in how each system processes bug reports, plans test strategies, or handles localization. The high number of resolved instances suggests that combining diverse methods, such as our pass-then-invert generation, multi-prompt ensembles, or the use of different models, could be more powerful than optimizing a single technique in isolation. As a result, future systems might benefit from hybrid approaches that leverage the complementary strengths of these tools.

\subsection{How does issue fix difficulty impact test generation?}

SWT-Bench-Verified includes difficulty annotations for each bug fix, categorized based on the estimated developer time to resolve the issue. We examine how our approach performs on bugs that are perceived to be difficult to fix because we believe reproducing these cases is especially valuable, as reproduction has been shown to significantly aid in resolving bugs \cite{beller2018dichotomy}. Table~\ref{tab:system-difficulty} shows total instances resolved by difficulty level across four BRT tools. \textsc{AssertFlip}\ performs competitively across all categories, resolving the most issues in the \textless 15 min and 15min--1hr ranges and even demonstrating competitive performance on more complex 1--4 hour bugs. These findings highlight that \textsc{AssertFlip}\ is not only effective on easy bugs but also performs well when tackling more complex issues.

\begin{table}[ht]
\centering
\small
\caption{Comparison of systems across difficulty categories}
\begin{tabular}{lccc}
\toprule
\textbf{Approach} & \textbf{<15 min fix} & \textbf{15 min - 1 hour} & \textbf{1--4 hours} \\
& \textbf{(172 total)} & \textbf{(225 total)} & \textbf{(36 total)} \\
\midrule
\rowcolor{gray!10}
\textbf{\textsc{AssertFlip}} & 90 & 92 & 7 \\
Amazon Q & 91 & 111 & 10 \\
Otter++ & 83 & 71 & 6 \\
OpenHands & 54 & 55 & 11 \\
\bottomrule
\end{tabular}
\label{tab:system-difficulty}
\end{table}

\subsection{How does issue clarity impact test generation?}

SWT-Bench-Verified is a curated dataset where professional developers manually reviewed each bug report. They removed any issues that were vague and underspecified. This means that each issue is clearly defined. However, real-world bug reports are often not so clean. Many are incomplete, poorly written, or confusing. To understand how this affects test generation, we look at what happens when we evaluate on SWT-Bench-Lite. SWT-Bench-Lite was designed as a smaller and more efficient version of SWT-Bench. It contains 276 tasks selected to reduce evaluation cost, be faster, and still cover a wide range of bugs. The authors of SWT-Bench-Lite employed automatic filtering to exclude issues with very short descriptions (fewer than 40 words) and multi-file edits. However, this filtering was not based on a manual review of issue quality. As a result, while SWT-Bench-Lite is intended to be easier to work with, it may still contain vague or unclear tickets.

Table~\ref{tab:lite-verified-breakdown} reports results on SWT-Bench-Lite, split by whether an issue also appears in SWT-Bench-Verified (overlap). We emphasize that this is not a direct comparison between the two datasets, rather it focuses on the Lite issues that are also present in Verified allowing us to isolate the effect of issue clarity under consistent task conditions. On the issues overlapping with the verified subset, \textsc{AssertFlip}\ reaches a 45.5\% F$\rightarrow$P success rate. On the Lite-only (potentially vague) issues, this drops to 31.4\%. This highlights how important clearly defined issue descriptions are for generating bug-revealing tests. Future work could involve handling vague or incomplete tickets, for example, by asking clarification questions or retrieving related context.

\begin{table}[ht]
\centering
\small
% \footnotesize
\caption{\textsc{AssertFlip}\ results on SWT-Bench-Lite split by whether the instance also appears in Verified.}
\begin{tabular}{lccc}
\toprule
\textbf{Subset} & \textbf{Instances} & \textbf{F$\rightarrow$P Total} & \textbf{F$\rightarrow$P Rate (\%)} \\
\midrule
Overlap with Verified & 88 & 40 & 45.5 \\
Lite-only & 188 & 59 & 31.4 \\
\midrule
\textbf{Total} & 276 & 99 & 36 \\
\bottomrule
\end{tabular}
\label{tab:lite-verified-breakdown}
\end{table}

\subsection{Does coverage matter?}

In our experiments, we noticed a consistent pattern across runs, whenever a bug was successfully resolved (F$\rightarrow$P), the test that triggered it usually has high coverage over the lines modified by the bug fix. On the other hand, when the bug was not resolved, the test coverage over the patch was usually low. This was true even when components of the pipeline changed. This suggests that coverage over the buggy lines could be a useful signal for predicting whether a test is likely to be valid. Table~\ref{tab:coverage-stats} shows $\Delta C$ coverage results for our pipeline obtained using the official evaluation and reporting scripts from SWT-Bench \cite{mundler2024swt}. The overall coverage delta across all instances is 49.1\%, but when we break it down by whether the bug was resolved, we see a clear separation at 78.4\% for resolved bugs vs. 26.1\% for unresolved ones. This suggests that higher coverage of buggy lines is associated with better outcomes. Our pipeline uses an LLM-based validator to filter out tests that fail for the wrong reasons. However, this validator is not perfect, and in some cases it rejects tests that are correct or accepts ones that are not. Thus, it is worth exploring whether we can improve the pipeline by combining LLM validation with coverage signals. One idea is to use coverage as a secondary filter. For example, we could reject any test that covers very few buggy lines. Another idea is to replace the LLM validation entirely with a threshold-based coverage check. However, coverage is only meaningful if the localization is accurate. When localization is noisy and the wrong files or lines are identified, a test with high coverage will not cover the correct code. This makes it harder to trust coverage as a signal in isolation. A promising direction is to combine both signals, use coverage to reject low-quality tests, and LLM validation to reason about correctness. We leave a full investigation of this idea to future work, but our early results suggest that coverage could be an effective component in filtering and validation.

\begin{table}[ht]
\centering
% \small
\caption{$\Delta$ Mean Change Coverage for \textsc{AssertFlip}\ on SWT-Bench-Verified.}
\begin{tabular}{lr}
\toprule
\textbf{Metric} & \textbf{Value (\%)} \\
\midrule
% F$\rightarrow$P Success Rate (S) & 43.6 \\
Coverage Delta (All) $\Delta^\text{all}$ & 49.1 \\
Coverage Delta (Resolved) $\Delta^{\mathcal{S}}$ & 78.4 \\
Coverage Delta (Unresolved) $\Delta^{not \mathcal{S}}$ & 26.1 \\
\bottomrule
\end{tabular}
\label{tab:coverage-stats}
\end{table}

\section{Threats to validity} \label{sec:threats-to-validity}
Our experiments rely on the SWT-Bench Lite and Verified datasets. These benchmarks are designed to evaluate BRTs, but they do not represent all types of bugs or all codebases. The test cases are derived from a small set of Python projects, which means the results may not apply to other languages or less common frameworks. While our approach is conceptually language-agnostic, extending it beyond Python would require replacing the test runner and adapting the prompt templates to the target language’s testing framework and syntax conventions.We have not yet evaluated it with other languages, and we therefore include this as a threat to validity.

A major threat is the unknown overlap between benchmark data and the training data of the LLMs we use. Models like GPT-4o are trained on data that is not publicly disclosed, so it is possible some of the benchmark code or bug reports were seen during training. We cannot fully control for this. This is a limitation for all prior work using closed LLMs \cite{mundler2024swt, ahmed2025otter, nashid2025issue2test, wang2024aegis, kang2023large}. We used GPT-4o for all test generation and validation in this study. While we acknowledge that the performance of \textsc{AssertFlip}\ may vary if a different LLM is used, this choice was made to keep the focus on the improvements derived from the approach itself, disregarding advances in LLMs themselves. Furthermore, this also lets us steer away from concerns of data leakage since the model has an early knowledge cutoff of October 2023. Prompting is a key part of our workflow. To reduce variation, we use a fixed prompt structure in all experiments and tune only a small set of parameters. All ablations keep prompts consistent except for the tested change. The prompts, regeneration limits, and run settings were chosen iteratively, not through exhaustive search. Bug localization is another key factor. We do not focus on localization itself in this study, and our main results utilize localization from Agentless \cite{xia2024agentless}. In real projects, localization may be less accurate. Further, since only code localization was available to us from past work, we do not perform test localization. Our design choices, however, allow us to create standalone tests that can be added to the test suite without concerns of modifying existing tests and can keep the BRTs well isolated. 

LLMs are inherently non-deterministic, and stochastic behavior is a known limitation across all LLM-based approaches. To minimize randomness, we set the sampling temperature to zero in all our experiments. Furthermore, we randomly selected ten instances from the full set using a uniform sampling procedure with a fixed seed, and ran each of those instances five times under the same configuration. This was done to provide transparency around expected variance while staying within our cost constraints. Five of the ten instances produced identical outcomes across all runs. Three instances were consistent in four out of five runs while the remaining two instances in three of the five runs. These results show that the tool behaves reliably across repeated executions with limited variation that is expected in systems driven by LLMs. Our analysis in the paper attempts to present a holistic picture of the effectiveness of our technique on established benchmarks and does not focus on the outcome of specific instances. This interpretation is consistent with prior work in this space.

\section{Conclusion}

In this paper, we introduce \textsc{AssertFlip}\ a novel approach for automated bug reproduction from issue reports. We evaluate \textsc{AssertFlip}\ on SWT-Bench and find that it outperforms all the known approaches on the leaderboard. \textsc{AssertFlip}\ achieves a 43.6\% success rate on the SWT-Bench-Verified subset, making it the most effective open tool for this task. This performance validates the strength of our Pass-then-Invert strategy. Future work could explore combining \textsc{AssertFlip}\ with complementary approaches to better leverage their strengths, look into handling incomplete or vague bug reports, and integrate coverage metrics into the validation process. \textsc{AssertFlip}\ is a promising step toward automating the bug reproduction process, helping enable faster and more efficient debugging and repair workflows.

\section{Data Availability}
Our code, prompts, and full experimental results are available at the following online repository: \url{https://github.com/uw-swag/AssertFlip}

\bibliographystyle{ACM-Reference-Format}
\bibliography{main}

\end{document}